\documentclass[letterpaper,onecolumn,prd,floats,preprintnumbers,showpacs]{revtex4}
\usepackage{graphicx}
\usepackage{amssymb}

\newcommand{\ca}{a}
\newcommand{\cb}{b}

\newcommand{\ogpre}{{\Omega_{\gamma_01}}}

\newcommand{\oaf}{{\Omega_{\ca1}}}

\newcommand{\obf}{{\Omega_{\cb1}}}

\newcommand{\ogf}{{\Omega_{\gamma_11}}}

\newcommand{\obs}{{\Omega_{\cb2}}}

\newcommand{\ogfs}{{\Omega_{\gamma_12}}}

\newcommand{\ogs}{{\Omega_{\gamma_22}}}

\newcommand{\fgf}{f_{\gamma_01}}
\newcommand{\faf}{f_{\ca1}}
\newcommand{\fbf}{f_{\cb1}}
\newcommand{\rf}{R_1}
\newcommand{\fgs}{f_{\gamma_12}}
\newcommand{\fbs}{f_{\cb2}}

\newcommand{\fgfexpr}{{\frac{4\ogpre}{4\ogpre + 3\oaf + 3\obf}}}
\newcommand{\fafexpr}{{\frac{3\oaf}{4\ogpre + 3\oaf + 3\obf}}}
\newcommand{\fbfexpr}{{\frac{3\obf}{4\ogpre + 3\oaf + 3\obf}}}
\newcommand{\rfexpr}{{\frac{4-\obf}{4-4\obf}}}
\newcommand{\fgsexpr}{{\frac{4\ogfs}{4\ogfs + 3\obs}}}
\newcommand{\fbsexpr}{{\frac{3\obs}{4\ogfs + 3\obs}}}

\newcommand{\ra}{r_{\ca}}
\newcommand{\rb}{r_{\cb}}
\newcommand{\fNL}{f_{\rm{NL}}}
\newcommand{\fNLa}{f_{\rm{NL}}^a}
\newcommand{\fNLb}{f_{\rm{NL}}^b}
\newcommand{\fNLsingle}{\fNL^{\rm single}}
\newcommand{\abs}[1]{\vert#1\vert}

\newcommand{\zgprel}{\zeta_{\gamma_0 (1)}}

\newcommand{\zgpren}{\zeta_{\gamma_0 (2)}}

\newcommand{\zgpre}{\zeta_{\gamma_0}}

\newcommand{\zgfl}{\zeta_{\gamma_1 (1)}}

\newcommand{\zgfn}{\zeta_{\gamma_1 (2)}}

\newcommand{\zgf}{\zeta_{\gamma_1}}

\newcommand{\zgsl}{\zeta_{\gamma_2 (1)}}

\newcommand{\zgsn}{\zeta_{\gamma_2 (2)}}

\newcommand{\zgs}{\zeta_{\gamma_2}}

\newcommand{\zal}{\zeta_{\ca (1)}}

\newcommand{\zbl}{\zeta_{\cb (1)}}

\newcommand{\zan}{\zeta_{\ca (2)}}

\newcommand{\zbn}{\zeta_{\cb (2)}}

\newcommand{\za}{\zeta_{\ca}}

\newcommand{\zb}{\zeta_{\cb}}

\newcommand{\zfl}{\zeta_{1 (1)}}

\newcommand{\zfn}{\zeta_{1 (2)}}

\newcommand{\zf}{\zeta_{1}}

\newcommand{\zsl}{\zeta_{2 (1)}}

\newcommand{\zsn}{\zeta_{2 (2)}}

\newcommand{\zs}{\zeta_{2}}

\newcommand{\zl}{\zeta_{(1)}}
\newcommand{\zn}{\zeta_{(2)}}

\newcommand{\half}{{\textstyle{\frac{1}{2}}}}

\newcommand{\commentj}[1]{}

\begin{document}

\title{Primordial non-Gaussianity from two curvaton decays}

\author{Hooshyar Assadullahi, Jussi V\"{a}liviita and David Wands}

\affiliation{Institute of Cosmology and Gravitation, University of
  Portsmouth, Portsmouth PO1 2EG, United Kingdom}

\begin{abstract}
We study a model where two scalar fields, that are subdominant
during inflation, decay into radiation some time after inflation has
ended but before primordial nucleosynthesis. Perturbations of these
two curvaton fields can be responsible for the primordial curvature
perturbation. We write down the full non-linear equations that
relate the primordial perturbation to the curvaton perturbations on
large scales, calculate the power spectrum of the primordial
perturbation, and finally go to second order to find the
non-linearity parameter, $\fNL$. We find large positive values of $\fNL$ if
the energy densities of the curvatons are sub-dominant when they
decay, as in the single curvaton case. But we also find a large
$\fNL$ even if the curvatons dominate the total energy density
in the case when the inhomogeneous radiation produced by the first curvaton decay is
diluted by the decay of a second nearly homogeneous curvaton. The
minimum value $\min(\fNL)=-5/4$ which we find is the same as in the
single-curvaton case.
\end{abstract}


\maketitle

\section{Introduction}

Theories beyond the standard model often contain a large number of scalar
fields in addition to the standard-model fields. In the very early universe it
is natural to expect the initial values of these fields to be displaced from
the minimum of their potential. If they are displaced by more than the Planck
scale then they can drive a period of inflation. But if they are
displaced from their minimum by less than the Planck scale they will oscillate
about the minimum of their potential once the Hubble rate, $H$, drops below
their effective mass. An oscillating massive field has the equation of state
(averaged over several oscillations) of a pressureless fluid. Thus the energy
density of a weakly interacting massive field tends to grow relative to
radiation in the early universe. Such fields must therefore decay before
the primordial nucleosynthesis era to avoid spoiling the standard, successful
hot big bang model.
And if the energy density of a late-decaying scalar is non-negligible when
it decays then any inhomogeneity in its energy density will be transfered
to the primordial radiation \cite{Mollerach:1989hu,LM97}. This is the curvaton
scenario for the origin of structure \cite{Enqvist:2001zp,LWcurvaton,Moroi:2001ct}.

A curvaton field, $\chi$, is supposed to have a negligible energy density during
inflation but once the Hubble rate drops below the curvaton mass after
inflation, the curvaton energy density grows relative to radiation, reaching
its maximum value, $\Omega_{\chi,{\rm decay}}$, just before the curvaton
decays.
If all the species are in thermal equilibrium
and the baryon asymmetry is generated after the curvaton
decays, then curvaton mechanism generates adiabatic density perturbations.
If not then the curvaton can leave a residual isocurvature perturbation
\cite{Bucher:1999re,Enqvist:2000hp}
correlated with the curvature perturbation \cite{Lyth:2002my}.
The amplitude of isocurvature modes are severely constrained by current data
\cite{Trotta:2006ww,Lewis:2006ma,Bean:2006qz,Keskitalo:2006qv}.
Thus if, for example, the baryon asymmetry is produced by the
out-of-equilibrium curvaton decay,
then we require $\Omega_{\chi,{\rm decay}} \sim 1$.

If we take seriously the multiplicity of scalar fields in the early
universe then we should consider models where more than one field
can contribute to the primordial density perturbation on large
scales \cite{Wands:2007bd}. Several authors have considered the
combination of perturbations from a curvaton field and the inflaton
field driving inflation
\cite{Bartolo:2002vf,Wands:2002bn,Ferrer:2004nv}. More recently Choi
and Gong \cite{Choi:2007fy} considered the primordial perturbations
that may result from multiple curvaton fields, showing that the
presence of more than one curvaton field affects the amplitude of
residual isocurvature perturbations and their correlation with the
curvature perturbation. In this paper we extend their analysis to
study the non-linear curvature perturbation and how the multiple
late-decaying scalar fields may affect the non-Gaussianity of the
primordial curvature perturbation. For simplicity we restrict our
analysis to the case of two curvatons, but it should be
straightforward to extend our analysis to three or more curvatons.

Deviations from an exactly Gaussian distribution of the primordial density perturbation is
conventionally given in terms of a non-linearity parameter, $\fNL$ \cite{KomatsuSpergel}.
The current upper bound from the
WMAP three-year data \cite{Spergel:2006hy} is $\abs{\fNL} < 114$
while Planck is expected to bring this down to $\abs{\fNL} < 5$
\cite{KomatsuSpergel}.
Galaxy cluster surveys can offer complementary constraints
\cite{Sefusatti:2006eu}.
Measurement of $\fNL$ would give a valuable test of
inflation. If primordial perturbations originate from fluctuations in a
canonical inflaton field, driving slow-roll inflation, then $\fNL$ is less
than unity \cite{Maldacena,Acquaviva:2002ud}.
However, if primordial perturbations
originate from fluctuations in a single curvaton field,
then $\fNL \sim 1/\Omega_{\chi,{\rm decay}}$
\cite{Lyth:2002my,Bartolo,LR,Valiviita:2006mz,Malik:2006pm,Sasaki:2006kq}.
Therefore $\fNL$ could thus be large if the curvaton does not dominate the
energy density of the universe when it decays.
In this paper we shall show that when we consider the decay of two curvaton
fields it is also possible that $\fNL$ is large even when the energy densities of
both curvaton fields are dominant when they decay.

This paper is organized as follows.
In Sec.~\ref{sect:prelim} we introduce a non-linear definition
of curvature perturbations, define the perturbation power spectrum and bispectrum, and
the non-linearity parameter $\fNL$ which describes the non-Gaussianity
of perturbations at leading order.
In Sec.~\ref{sect:nonlineqns} we write down the full non-linear equations
that relate the curvaton perturbations to the total and radiation perturbations
at the first-curvaton decay and at the second-curvaton decay, and finally to the primordial
curvature perturbation. Then we solve these equations up to the second order and
find the resulting $\fNL$ of the primordial perturbation before nucleosynthesis.
The general expression for $\fNL$ is a non-trivial function of four parameters.
To get an insight into its behavior we analyze some special cases
in Sec.~\ref{sect:variouslimits}. Finally, in Sec.~\ref{sect:conc} we
present concluding remarks. As the notation becomes rather heavy in the two-curvaton scenario,
we list most of the symbols used in this paper in Appendix~\ref{sect:appendixa}.


\section{Preliminaries}
\label{sect:prelim}

\subsection{Primordial curvature perturbation}

The primordial density perturbation can be described in terms of the
non-linearly perturbed expansion on uniform-density hypersurfaces
\cite{LMS} (see also \cite{Rigopoulos:2003ak,Langlois:2005qp})
\begin{equation}
 \label{eqn:zetanl}
\zeta (t,{\bf x}) = \delta N (t,{\bf x}) +
 \frac13 \int_{\bar\rho(t)}^{\rho(t,{\bf x})}
 \frac{d\tilde\rho}{\tilde\rho+\tilde{p}} \,,
\end{equation}
where $N=\int Hdt$ is the integrated local expansion, $\tilde\rho$ the
local density and $\tilde{p}$ the local pressure, and $\bar\rho$ is the homogeneous density in the background model.

We will expand the curvature perturbation at each order $(n)$ as
\begin{equation}
\zeta(t,{\bf x}) = \sum_{n=1}^{\infty}\frac{1}{n!}\zeta_{(n)}(t, {\bf x})\,,
\label{eqn:expansion}
\end{equation}
where we assume that the first-order perturbation, $\zeta_{(1)}$, is Gaussian as it is
proportional to the initial Gaussian field perturbations.
Higher-order terms describe the non-Gaussianity of the full non-linear $\zeta$.

Working in terms of the Fourier transform of $\zeta$, we define the primordial power spectrum as
\begin{equation}
\langle \zeta({\bf k_1}) \zeta({\bf k_2}) \rangle = (2\pi)^3
P_\zeta(k_1) \delta^3({\bf k_1}+{\bf k_2}) \,.
\end{equation}
The average power per logarithmic interval in Fourier space is given by
\begin{equation}
 {\cal P}_\zeta(k) = \frac{4\pi k^3}{(2\pi)^3} P_\zeta(k) \,,
\end{equation}
and is roughly independent of wavenumber $k$.

The primordial bispectrum is given by
\begin{equation}
 \label{defB}
\langle \zeta({\bf k_1})\zeta({\bf k_2})
\zeta({\bf k_3}) \rangle
 =  (2\pi)^3 B(k_1,k_2,k_3) \delta^3({\bf k_1}+{\bf
k_2}+{\bf k_3}) \,.
\end{equation}
The bispectrum vanishes for a purely Gaussian distribution, and hence is non-zero only at fourth and higher-order.
The amplitude of the bispectrum relative to the power spectrum is commonly parameterized in terms of the non-linearity parameter, $\fNL$, defined such that \cite{KomatsuSpergel}
\begin{eqnarray}
B(k_1,k_2,k_3) \!\! & = & \!\! (6/5)\fNL \left[ P(k_1) P(k_2) + 2\,{\rm perms} \right]\,.
\label{fnl}
\end{eqnarray}
Higher order statistics, like trispectrum
(see e.g. \cite{Sasaki:2006kq,Byrnes:2006vq,Byrnes:2007tm}) or full non-linear probability density
function of the primordial $\zeta$  \cite{Sasaki:2006kq}, can give also valuable information
on non-Gaussianity in the curvaton model, but in this paper we consider the bispectrum only.

\subsection{Curvaton perturbations}

We will consider the primordial curvature perturbation produced by the decay of two scalar fields $a$ and $b$.
Without loss of generality, we assume that the curvaton $a$ decays first when $H=\Gamma_a$ followed by the decay of the curvaton $b$ when $H=\Gamma_b$, where $\Gamma_b<\Gamma_a$.

Small-scale (sub-Hubble) vacuum fluctuations of any light scalar field are stretched by the expansion to large (super-Hubble) scales during a period of inflation in the early universe.
If the curvaton fields are weakly-coupled, massive scalar fields whose masses are less than the
Hubble scale, $H_\ast \gg m$, during inflation, then they acquire an almost
scale-invariant spectrum of perturbations on super-Hubble scales,
\begin{equation}
 {\cal P}_{\delta\ca_{\ast}} = {\cal P}_{\delta\cb_{\ast}} = \left( \frac{H_\ast}{2\pi} \right)^2 \,,
\end{equation}
where $H_\ast$ is the Hubble rate at Hubble exit.
In Refs.~\cite{Seery:2005gb,Seery:2006js,Seery:2006vu,Yokoyama:2007uu,Sasaki:2007ay}
it has been shown that in general,
if slow-roll conditions are satisfied, the non-Gaussianity
of field perturbations at Hubble exit is small
(at least for the three-point and four-point correlators). Consequently,
in what follows, we assume that the field perturbations at Hubble exit have
Gaussian and independent distributions, consistent with weakly-coupled isocurvature
fluctuations of light scalar fields \cite{Enqvist:2004bk,Vaihkonen:2005hk}.

Once the Hubble rate drops below the mass of each curvaton field the field begins to oscillate.
The local value of each curvaton can evolve between Hubble-exit during inflation
and the beginning of the field oscillations.
We parameterize this evolution by functions $g_{\ca}$ and $g_{\cb}$, but we
assume the two fields
remain decoupled so that their perturbations remain uncorrelated. Thus at the
beginning of curvaton oscillations, the curvaton fields have values
\begin{eqnarray}
\label{defga}
\ca_{\rm osc} & = & g_{\ca}(\ca_\ast)\\
\label{defgb}
\cb_{\rm osc} & = & g_{\cb}(\cb_\ast)\,.
\end{eqnarray}

We can define non-linear
perturbation for each curvaton
analogous to the total
perturbation (\ref{eqn:zetanl})
\begin{equation}
 \label{eqn:zetanlcurvaton}
\za
  = \delta N + \frac13 \ln \left( \frac{\rho_\ca}{\bar\rho_\ca} \right)
  = \delta N + \frac23 \ln \left( \frac{\ca_{\rm osc}}{\bar{\ca}_{\rm osc}} \right)
  \,.
\end{equation}
where we take the energy density to be proportional to the square of the field value,
$\rho_a \propto \ca_{\rm osc}^2$, when the field begins to oscillate, and similarly for $\zb$.
On uniform-curvaton-density hypersurfaces \cite{Lyth:2003im} $\za$ and $\zb$ are curvature perturbations.
The curvature perturbation $\za$ (and $\zb$) becomes constant on scales larger than the Hubble scale once
each curvaton starts oscillating, and while we can neglect energy transfer to the radiation,
$\Gamma_\ca<H$.
Thus we require
\begin{equation}
 \Gamma_\ca \ll m_\ca \quad {\rm and} \quad  \Gamma_\cb \ll m_\cb \,,
\end{equation}
consistent with the assumption that we are dealing with weakly coupled fields.
For simplicity we will further assume that $\Gamma_\ca\ll m_\cb$ and that the energy density of both curvatons is negligible when they begin oscillating, which ensures that we can neglect any gravitational coupling between fields in the overdamped regime before they begin oscillating.

At first order the curvature perturbation (\ref{eqn:zetanlcurvaton}) for each curvaton can thus be given in terms of the field perturbations on spatially flat hypersurfaces ($\delta N=0$)
\begin{eqnarray}
\zal & = & \frac{2}{3} \left.\frac{\delta\ca}{\bar{\ca}}\right|_{\rm osc} =
\frac{2}{3}\frac{g'_{\ca}}{g_\ca}\delta\ca_{\ast}\,,\label{eqn:zal}\\
\zbl & = & \frac{2}{3} \left.\frac{\delta\cb}{\bar{\cb}}\right|_{\rm osc} = \frac{2}{3}\frac{g'_{\cb}}{g_\cb}\delta\cb_{\ast}\,,\label{eqn:zbl}
\end{eqnarray}
where we have Taylor expanded the functions $g_{\ca}$ and $g_{\cb}$, defining
$g'_{\ca}\equiv \partial g_{\ca} / \partial\ca_{\ast}$, and
$g'_{\cb}\equiv \partial g_{\cb} / \partial\cb_{\ast}$.

The power spectra $P_{\za}$ and $P_{\zal}$ are the same at leading order. The higher order
corrections are generally so small that we use $P_{\za}$ and $P_{\zal}$ interchangeably in the following.
Therefore the power spectra of $\za$ and $\zb$ when the curvatons start to oscillate
are related by
\begin{equation}
P_{\zb} = \beta^2 P_{\za}\,,
\label{eqn:PbvsPa}
\end{equation}
where
\begin{equation}
\beta = \frac{g'_{\cb}/g_{\cb}}{g'_{\ca}/g_{\ca}}\,.
\end{equation}
If we assume linear evolution between Hubble exit and the beginning
of curvaton oscillations, this factor reduces to the ratio
of the background curvaton field values at Hubble exit,
\begin{equation}
 \beta = \frac{\ca_{\ast}}{\cb_{\ast}}\,.
\end{equation}

At second order Eq. (\ref{eqn:zetanlcurvaton}) gives on spatially flat hypersurfaces
($\delta N=0$)
\begin{eqnarray}
  \zan & = & -\frac{3}{2}\left( 1 - \frac{g_{\ca} g''_{\ca}}{g'^2_{\ca}}
  \right) \zal^2\,,\label{eqn:zan}\\
  \zbn & = & -\frac{3}{2}\left( 1 - \frac{g_{\cb} g''_{\cb}}{g'^2_{\cb}}
  \right) \zbl^2\,,\label{eqn:zbn}
\end{eqnarray}
where $\zal$ and $\zbl$ are given in (\ref{eqn:zal}) and (\ref{eqn:zbl}).
Analogous to (\ref{fnl}) we can define the non-linearity parameters of each
curvaton $\fNLa$ and $\fNLb$. Then comparing to (\ref{eqn:zan}) and (\ref{eqn:zbn}) we find
\begin{eqnarray}
\fNLa & = & -\frac{5}{4}\left( 1 - \frac{g_{\ca} g''_{\ca}}{g'^2_{\ca}}
\right)\,,\\
\fNLb & = &  -\frac{5}{4}\left( 1 - \frac{g_{\cb} g''_{\cb}}{g'^2_{\cb}}
  \right)\,.
\end{eqnarray}
If the evolution of field values is linear between Hubble exit and the beginning
of curvaton oscillations, we have $\fNLa = \fNLb = -5/4$.
Although it is possible to construct potentials which lead to a non-linear
evolution \cite{nurmi}, we assume henceforth in this paper that
$g''_{\ca} = g''_{\cb} = 0$, consistent with weakly interacting fields.

\subsection{Calculating $\fNL$}

Before calculating how the two curvaton perturbations contribute to the primordial curvature perturbation in detail, we can first consider the general form of the primordial power spectrum and bispectrum that will result from a generic model where the local values of two Gaussian fields determine the primordial perturbation.

The primordial curvature perturbation after both curvatons decay can be
written up to second order in terms of the first order curvaton perturbations
as
\begin{eqnarray}
\zeta \equiv \zs & = & \zsl + \half \zsn\nonumber\\
& = & A\zal + B\zbl + \half C \zal^2 + \half D\zbl^2\ + \half
E\zal\zbl,
 \label{eqn:zetageneral}
\end{eqnarray}
where $\zs$ is the total curvature perturbation after the
second decay. The first order part $\zsl$ is Gaussian, since
it is a linear function of the field perturbations at Hubble exit (and for weakly
coupled light scalar fields these field perturbations are Gaussian).
The second order part $\zsn$ is non-Gaussian, since it contains squares of Gaussian
variables.
$A$, $B$, $C$, $D$ and $E$ are coefficients that will depend on
background quantities at the time of the first
and second decay. In Sect.~\ref{sect:nonlineqns} we derive the
expressions for them in the two-curvaton decay scenario.

The primordial power spectrum at leading order is
\begin{equation}
   \label{Pzetageneral}
   P_\zeta
   = A^2 P_{\za} +  B^2 P_{\zb}\,.
\end{equation}
Employing the relation (\ref{eqn:PbvsPa}) we can write
\begin{eqnarray}
P_{\za} & = & \frac{1}{A^{2}+\beta^2 B^{2}}P_{\zeta}
\,,\label{Pzabofz}\\
P_{\zb} & = & \frac{\beta^2}{A^{2}+\beta^2 B^{2}}P_{\zeta}
\,.
 \label{Pzbbofz}
\end{eqnarray}

Since the two curvatons are uncorrelated with one another, i.e.,
\begin{equation}
\big\langle \zal({\bf k_{1}}) \zbl({\bf k_{2}}) \big\rangle = 0
\end{equation}
for any wave vectors ${\bf k_{1}}$ and ${\bf k_{2}}$,
we find that at leading order the three-point correlator of the primordial
perturbation is
\begin{eqnarray}
&&\Big\langle  \zeta({\bf k_{1}}) \zeta({\bf k_{2}})\zeta({\bf k_{3}}) \Big\rangle = \nonumber\\
&& \;\Big[\half A^2 C
\Big\langle  \zal({\bf k_{1}}) \zal({\bf k_{2}})(\zal\!\ast\!\zal)({\bf k_{3}}) \Big\rangle
 + \half B^2 D
\Big\langle  \zbl({\bf k_{1}}) \zbl({\bf k_{2}})(\zbl\!\ast\!\zbl)({\bf k_{3}}) \Big\rangle \nonumber\\
&& +\half A B E
\Big\langle  \zal({\bf k_{1}}) \zbl({\bf k_{2}}) (\zal\!\ast\!\zbl)({\bf k_{3}}) \Big\rangle
\Big] + \mbox{ 2 permutations of } \{ {\bf k_1, k_2, k_3} \}\,,
\label{eqn:3correlator1}
\end{eqnarray}
where $\ast$ denotes a convolution. For example,
\begin{equation}
(\zal\!\ast\!\zal)({\bf k_{3}})= \frac{1}{(2\pi)^3}
 \int d^{3}{\bf q} \zal({\bf q}) \zal({\bf k_{3}}-{\bf q})\,.
\end{equation}
%
Note that there are two ways to form two pairs of correlators in the first two terms on the right-hand-side of (\ref{eqn:3correlator1}).
Therefore, the three-point correlator simplifies to
\begin{eqnarray}
\langle  \zeta({\bf k_{1}}) \zeta({\bf k_{2}})\zeta({\bf k_{3}}) \rangle & = &
\Big\{ \big[ A^2 C  P_{\zal}(k_1)
P_{\zal}(k_2) +  B^2
  D P_{\zbl}(k_1)
P_{\zbl}(k_2)\nonumber\\
&& + \half A B E  P_{\zal}(k_1)
P_{\zbl}(k_2) \big]
+ \mbox{ 2 perms} \Big\}\times (2\pi)^3\delta^{(3)}({\bf k_1 + k_2 + k_3})\,.
\label{eqn:3correlator2}
\end{eqnarray}

Substituting (\ref{Pzabofz}) and~(\ref{Pzbbofz}) into
(\ref{eqn:3correlator2}) leads to
\begin{eqnarray}
\langle  \zeta({\bf k_{1}}) \zeta({\bf k_{2}})\zeta({\bf k_{3}}) \rangle & = &
\frac{A^2 C
 + \beta^4  B^2 D + \half \beta^2 A B E}{\left(A^{2}+\beta^2 B^{2}\right)^2}
\nonumber\\
&& \left\{ P_{\zeta}(k_1) P_{\zeta}(k_2) + \mbox{ 2 perms} \right\}
 \times (2\pi)^3\delta^{(3)}({\bf k_1 + k_2 + k_3})\,.
\label{eqn:3correlator3}
\end{eqnarray}
Hence the bispectrum, defined in Eq.~(\ref{defB}), is in the form given by Eq.~(\ref{fnl}), where the non-linearity parameter is
\begin{equation}
\fNL = \frac{5}{6}
\frac{ C A^2 + \half \beta^2 E A B + \beta^4 D B^2}
     {\left(A^{2}+\beta^2 B^{2}\right)^2}\,.
\label{eqn:fnl}
\end{equation}

\section{Full non-linear equations}
\label{sect:nonlineqns}

We will estimate the primordial density perturbation produced by the
decay of two curvaton fields some time after inflation has ended
using the sudden-decay approximation \cite{Lyth:2002my},
generalizing the non-linear analysis of Ref.~\cite{Sasaki:2006kq} to
the case of two curvatons. In this approximation the curvaton fields
and the radiation are treated as non-interacting fluids except at
the instant of decay when all the energy density of the curvaton is
transfered to radiation. This should be a good approximation for
scales much larger than the decay time, $\Gamma^{-1}$, and has been
shown to give a good estimate of the primordial non-Gaussianity when
compared against numerical simulations
\cite{Valiviita:2006mz,Malik:2006pm,Sasaki:2006kq}.

Before either curvaton has decayed, the radiation (from inflaton
decay products) has a definite equation of state,
$p_{\gamma}=(1/3)\rho_{\gamma}$, and so do the oscillating curvaton
fields, $p_a=p_b=0$. Thus we have three non-interacting fluids with
barotropic equations of state and hence three curvature
perturbations (\ref{eqn:zetanl}) which are constant on large scales
\cite{Lyth:2003im}
\begin{eqnarray}
 \zeta_\gamma &=& \delta N + \frac14 \ln \left(\frac{\rho_{\gamma}}{\bar\rho_{\gamma}} \right) \,,\label{eqn:zgfullnl}\\
 \za &=& \delta N + \frac13 \ln \left( \frac{\rho_a}{\bar\rho_a} \right) \,,\label{eqn:zafullnl}\\
 \zb &=& \delta N + \frac13 \ln \left( \frac{\rho_b}{\bar\rho_b} \right) \,\label{eqn:zbfullnl}.
\end{eqnarray}

On the spatial hypersurface where $H=\Gamma_a$, there is an abrupt
jump in the overall equation of state due to the sudden decay of the
first curvaton into radiation, but the total energy density is
continuous
\begin{equation}
 \rho_{\gamma_11} + \rho_{\cb1} = \rho_{\gamma_01} + \rho_{\ca1} + \rho_{\cb1}\,.
\label{eqn:totrhoat1}
\end{equation}
Here $\rho_{\gamma_11}$ is the radiation energy density immediately after the
first curvaton decay, $\rho_{\cb1}$ is density of the second curvaton at the time
of the first curvaton decay, $\rho_{\gamma_01}$ is the density of
pre-existing radiation just before the first curvaton decay, and $\rho_{\ca1}$ is
the density of the first curvaton just before the first decay when it is converted
to radiation. Eq.~(\ref{eqn:totrhoat1}) simplifies to
\begin{equation}
 \rho_{\gamma_11} = \rho_{\gamma_01} + \rho_{\ca1}\,.
\label{eqn:rhogat1}
\end{equation}

Note that the first-decay hypersurface is a uniform-density hypersurface and
thus, from Eq.~(\ref{eqn:zetanl}) the perturbed expansion on this hypersurface
is $\delta N=\zf$, where $\zeta_1$ denotes the total curvature perturbation
at the first-decay hypersurface.
Hence on the first-decay hypersurface we have
 \begin{eqnarray}
 \rho_{\gamma_01} & = & \bar\rho_{\gamma_01} e^{4(\zgpre-\zf)} \,,\label{eqn:rhog0}\\
 \rho_{\gamma_11} & = & \bar\rho_{\gamma_11} e^{4(\zgf-\zf)} \,,\label{eqn:rhog1}\\
 \rho_{\ca1} & = & \bar\rho_{\ca1} e^{3(\za-\zf)} \,,\label{eqn:rhoa1}\\
 \rho_{\cb1} & = & \bar\rho_{\cb1} e^{3(\zb-\zf)} \,,\label{eqn:rhob1}
\end{eqnarray}
where we employed Eqs.~(\ref{eqn:zgfullnl}--\ref{eqn:zbfullnl}).
In the curvaton scenario we usually assume the initial radiation is
homogeneous before the curvaton decays, so that $\zgpre=0$, but in this
section we keep $\zgpre$ in our calculations for generality.
On a uniform-density hypersurface the total energy density
is homogeneous. Therefore an infinitesimal time before the first decay
we have
\begin{equation}
\rho_{\gamma_01}+\rho_{\ca1}+\rho_{\cb1} = \bar\rho_1.
\end{equation}
Substituting here Eqs.~(\ref{eqn:rhog0}), (\ref{eqn:rhoa1}) and
(\ref{eqn:rhob1}), and dividing by the total energy density $\bar\rho_1$
we end up with
\begin{equation}
  \ogpre e^{4(\zgpre - \zf)} + \oaf e^{3(\za - \zf)} + \obf e^{3(\zb - \zf)} =
  1\,,
\label{eqn:nonlinf1}
\end{equation}
where $\ogpre = \rho_{\gamma_01}/\bar\rho_1$, $\oaf=\rho_{\ca1}/\bar\rho_1$,
and $\obf=\rho_{\cb1}/\bar\rho_1$ are the energy density parameters
of the pre-existing radiation, the first curvaton, and the second curvaton at
the first curvaton decay, respectively.
On the other hand, an infinitesimal time after the first curvaton has
decayed into radiation the total energy density can be written as
\begin{equation}
\rho_{\gamma_11}+\rho_{\cb1} = \bar\rho_1\,,
\end{equation}
which gives an equation
\begin{equation}
  \ogf e^{4(\zgf - \zf)} + \obf e^{3(\zb - \zf)} = 1\,.
\label{eqn:nonlinf2}
\end{equation}
The total curvature perturbation $\zeta$ is the perturbation on the decay hypersurface and thus is continuous between the two phases. However the energy density of radiation changes abruptly at the decay time and the radiation curvature perturbation is discontinuous, $\zgpre\neq\zgf$.

Similarly, an infinitesimal time before the second decay we have
\begin{equation}
  \ogfs e^{4(\zgf - \zs)} + \obs e^{3(\zb - \zs)} = 1\,.
\label{eqn:nonlins}
\end{equation}
while an infinitesimal time after the second decay we have
\begin{equation}
  \ogs e^{4(\zgs - \zs)} = 1\,.
\label{eqn:nonlinfinal}
\end{equation}
Above $\ogfs$ and $\obs$ are the radiation and the second curvaton
energy density parameters just before the second curvaton decay,
$\ogs$ is the radiation density parameter just after the second
curvaton decay, and $\zs$ is the total curvature perturbation at the
second-curvaton decay hypersurface. As the radiation is the only
constituent left after the second decay, the radiation curvature
perturbation $\zgs$ equals the the total curvature perturbation
$\zs$, and Eq.~(\ref{eqn:nonlinfinal}) reduces to $\ogs = 1$ as we
will see in the next subsection. After both curvatons have decayed we
have a single radiation fluid with equation of state $p=\rho/3$ and
hence the curvature perturbation remains constant. Thus we identify
the primordial curvature perturbation, 
\begin{equation}
\zeta=\zgs.
\label{eqn:zetazeta}
\end{equation}

We can Taylor expand the exponential functions, $e^x = 1 + x + x^2/2 +
\ldots$, in Eqs.~(\ref{eqn:nonlinf1}--\ref{eqn:nonlinfinal}) to express the
final curvature perturbation, $\zeta$, in terms of the initial curvaton
perturbations, $\za$ and $\zb$, at any given order.

\subsection{Zeroth order}

Taylor expanding Eq.~(\ref{eqn:nonlinf1}) at zeroth order gives
\begin{equation}
\ogpre + \oaf + \obf = 1\,.
\end{equation}
This is identically true, since we are studying spatially flat models with $\Omega_{\rm tot} =1$.
Similarly Eq.~(\ref{eqn:nonlinf2}) gives
\begin{equation}
\ogf + \obf = 1\,.
\end{equation}
Therefore $\ogf$ is a redundant variable, and in what follows
we can replace it by $1-\obf$. To zeroth order, equation (\ref{eqn:nonlins})
is
\begin{equation}
\ogfs + \obs = 1\,.
\end{equation}
Therefore $\ogfs$ is  a redundant variable, and in what follows
we can replace it by $1-\obs$. Finally Eq.~(\ref{eqn:nonlinfinal}) is
identically true as it states $\ogs = 1$.
At all higher orders
Eq.~(\ref{eqn:nonlinfinal}) states $\zgs=\zs$.
After the second decay only radiation is left.

\subsection{First order}

\subsubsection{First decay}

At first order, Eq.~(\ref{eqn:nonlinf1}) gives
\begin{equation}
  4 \ogpre [\zgprel - \zfl] + 3 \oaf [\zal - \zfl] + 3 \obf [\zbl - \zfl] = 0\,.
\end{equation}
{}From this we solve for the total curvature perturbation, $\zfl$,
at the first decay surface. It is
\begin{equation}
  \zfl = \fgf\zgprel + \faf\zal + \fbf\zbl\,,
\label{eqn:zfl}
\end{equation}
where
\begin{eqnarray}
\fgf & = & \fgfexpr\,,\\
\faf & = & \fafexpr\,,\\
\fbf & = & \fbfexpr\,.
\end{eqnarray}
These first order curvature transfer efficiency parameters at the first decay
obey the relation $\fgf + \faf + \fbf = 1$. In what follows we use this to
eliminate $\fgf$.

After the first curvaton decays, but before the second curvaton decays, the curvature perturbation in the radiation, $\zgf$, will remain constant on large scales.
Eq.~(\ref{eqn:nonlinf2}) relates the radiation curvature perturbation and the
total curvature perturbation immediately after the first decay. At first order it reads
 \begin{equation}
  4\ogf [\zgfl - \zfl] + 3\obf [\zbl - \zfl] = 0\,,
\end{equation}
which gives
\begin{equation}
  \zgfl = \rf\zfl - \left(\rf-1\right)\zbl\,,
\label{eqn:zgfl1}
\end{equation}
where
\begin{eqnarray}
\rf & = & \rfexpr\\
& =& \frac{3+\faf}{3(1-\fbf)+\faf}\,.
\label{eqn:rf}
\end{eqnarray}
Note that if the density of the second curvaton is negligible when the first curvaton decays then we have $R_1=1$ and $\zgfl=\zfl$.

Finally, we substitute (\ref{eqn:zfl}) into (\ref{eqn:zgfl1})
\begin{equation}
\zgfl = \rf\left(1-\faf-\fbf\right)\zgprel + \rf\faf\zal + \left[ 1 - \rf\left( 1-\fbf \right)\right]\zbl\,.
\label{eqn:zgfl2}
\end{equation}
This remains constant between the first and second decay and hence, at the second decay
we can use this as the incoming radiation perturbation.

\subsubsection{Second decay}

Following the same procedure as above, we find from Eq.~(\ref{eqn:nonlins}) at first order
(or equivalently using (\ref{eqn:zfl}) but dropping curvaton $\cb$, then relabeling
$\ca\rightarrow \cb$, $1\rightarrow 2$ and $0\rightarrow1$)
\begin{equation}
\zsl =  \fgs\zgfl + \fbs\zbl\,,
\label{eqn:zsl1step}
\end{equation}
where
\begin{eqnarray}
\fgs & = & \fgsexpr\,,\\
\fbs & = & \fbsexpr\,.
\end{eqnarray}
These first order curvature transfer efficiency parameters at second decay
obey the relation $\fgs + \fbs = 1$,
and hence
Eq.~(\ref{eqn:zsl1step}) reads
\begin{equation}
\zsl = \left( 1 - \fbs \right) \zgfl + \fbs\zbl\,.\\
\end{equation}
Finally, using Eq.~(\ref{eqn:zgfl2}) for $\zgfl$ we obtain
\begin{eqnarray}
\zsl
& = & \rf\left(1-\faf-\fbf \right)\left( 1 - \fbs \right)\zgprel + \ra\zal +
\rb\zbl\,,
\label{eqn:zgslwithrad}
\end{eqnarray}
where
\begin{eqnarray}
\ra & = & \rf\faf\left( 1 - \fbs \right)\\
    & = & \frac{(1-\fbs)(3+\faf)\faf}{3(1-\fbf)+\faf}\,,\label{eqn:faformula}\\
\rb & = & 1 - \rf \left(1-\fbf\right) \left(1-\fbs\right)\\
    & = & \frac{(1-\fbf)\fbs(3+\faf)+\fbf\faf}{3(1-\fbf)+\faf}\,\label{eqn:fbformula}.
\end{eqnarray}

From Eqs.~(\ref{eqn:nonlinfinal}) and (\ref{eqn:zetazeta})
we have $\zsl=\zgsl=\zl$. Therefore, if the pre-existing radiation perturbation vanishes, $\zgprel = 0$,
the primordial first order curvature perturbation after the second decay is
\begin{equation}
  \zl = \ra\zal + \rb\zbl\,.
\label{eqn:zgsl}
\end{equation}
Then in Eq.~(\ref{eqn:zetageneral}) at first order we can identify
\begin{equation}
 A=\ra,\ \ \ \ \mbox{ and } \ B=\rb\,,
\label{eqn:AandB}
\end{equation}
and the power spectrum for the primordial curvature perturbation is
given by Eq.~(\ref{Pzetageneral}).
\begin{equation}
\frac{P_{\zl}}{P_{\zal}} = {\ra^2+\beta^2\rb^2}\,.
\label{eqn:transfeffgeneral}
\end{equation}
where $\beta^2$ defined in Eq.~(\ref{eqn:PbvsPa}) gives the ratio between the initial power in the curvaton $\ca$ and that in the curvaton $\cb$.

\begin{figure*}[!ht]
\includegraphics[width=0.72\textwidth]{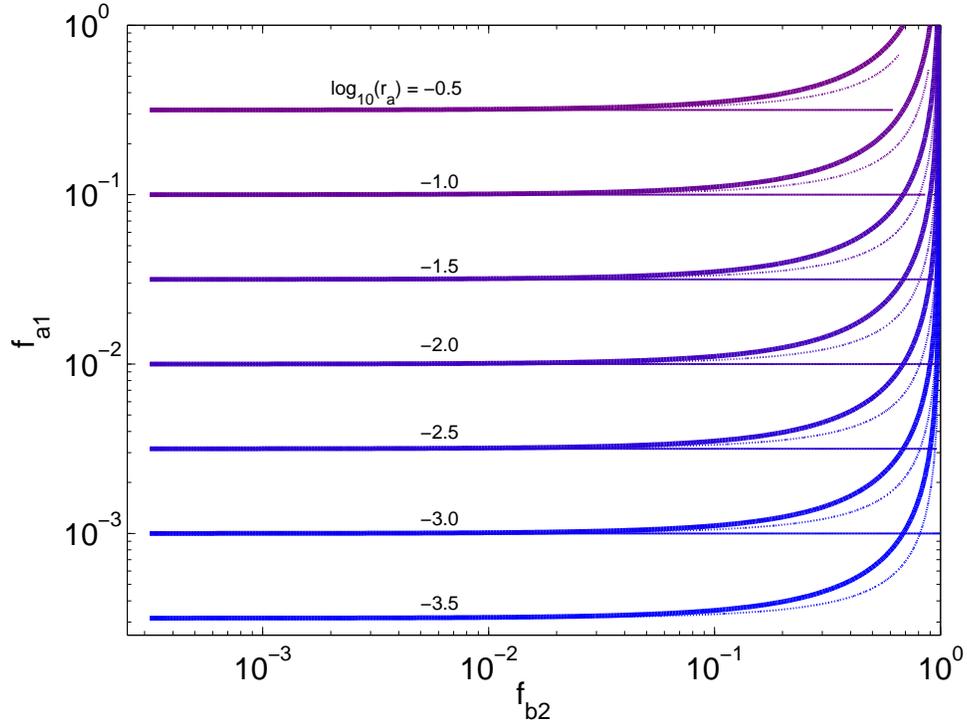}
\caption{$\log_{10}\ra(\fbs,\faf)$ for $\fbf=0$ (\emph{thick solid lines}), for $\fbf=\fbs/2$
  (\emph{dotted lines}) and for $\fbf=(1+\faf/3)\fbs$ (\emph{thin solid lines}). In the latter two cases
   the constraint (\ref{eqn:rbfconstraint1}) excludes a small
  region in the top right corner. 
  The contours of equal $\ra$ are shown, from bottom to top, for
  $\ra=$ $10^{-3.5}$, $10^{-3}$, $10^{-2.5}$, $10^{-2}$, $10^{-1.5}$,
  $10^{-1}$, and $10^{-0.5}$. \label{fig:ra}}
\end{figure*}

\begin{figure*}[!ht]
\includegraphics[width=0.72\textwidth]{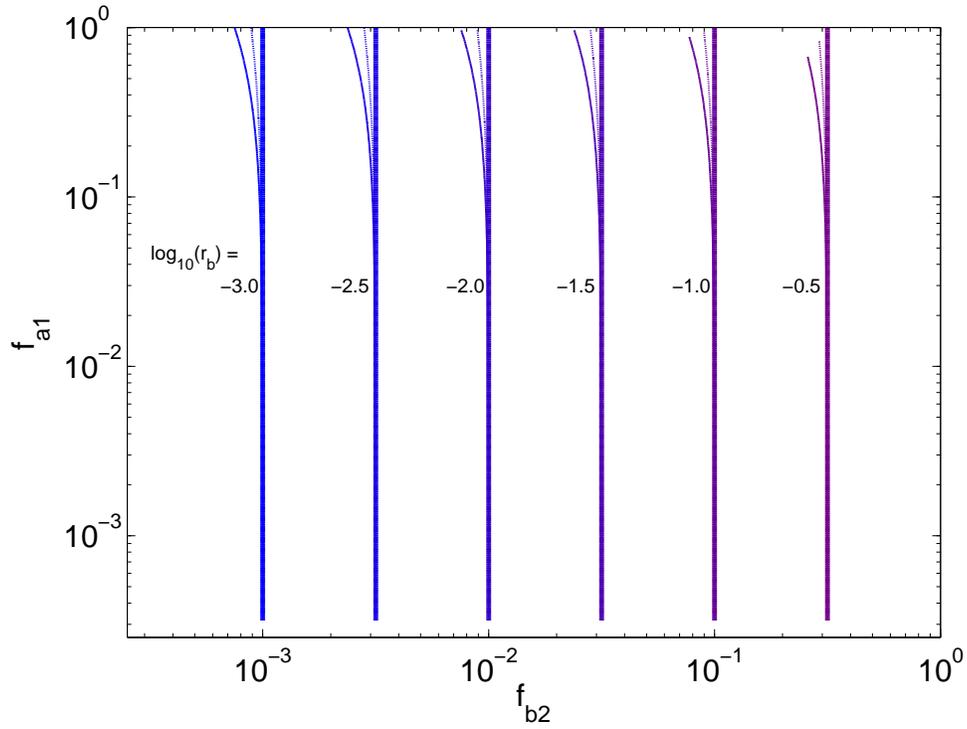}
\caption{Same as Fig.~\ref{fig:ra} but now for $\log_{10}\rb$.
  The contours of equal $\rb$ are shown, from left to right, for
$\rb=$ $10^{-3}$, $10^{-2.5}$, $10^{-2}$, $10^{-1.5}$,
$10^{-1}$, and $10^{-0.5}$. \label{fig:rb}}
\end{figure*}

Note that our result differs from that presented recently by Choi and Gong \cite{Choi:2007fy} due to the presence of the extra factor $\rf$ which arises due to the difference between the uniform total density hypersurface and the uniform radiation density hypersurface when curvaton $\ca$ decays, if the density of the curvaton $\cb$ is not negligible. When $\obf=0$ we find from Eq.~(\ref{eqn:rf}) that $\rf=1$ and we recover the simpler result \cite{Choi:2007fy}
\begin{equation}
 \label{Choilinear}
 \zgsl = (1-\fbs)\faf\zal+\fbs\zbl \,.
\end{equation}
The transfer coefficients $\ra$ and $\rb$ in this case are shown by the thick solid lines in Figs.~\ref{fig:ra} and
\ref{fig:rb}.

However in general when $\obf\neq0$ and $\rf\neq1$ we find a novel
effect where the inhomogeneous density of the curvaton $\cb$ may
lead to a perturbation in the radiation density after the curvaton
$\ca$ decays, even if the curvaton $\ca$ is homogeneous, see
Eq.~(\ref{eqn:zgfl2}). A local overdensity of the curvaton $\cb$
delays the decay of curvaton $\ca$ due to a local gravitational time
dilation, resulting in a local overdensity in the radiation after
decay. In practice $\fbs\geq(3/4)\fbf$ so this is usually a small
correction. The dotted lines in Figs.~\ref{fig:ra} and
\ref{fig:rb} show this effect for the case $\fbf = \fbs/2$.


When considering $\ra$ and $\rb$ it should be noted that the
range of $\faf$, $\fbf$, and $\fbs$ is $[0,\,1]$, but
the allowed range of $\fbf$ is constrained for given $\faf$ and $\fbs$.
Since $\faf + \fbf = 1 - \fgf$, and $\fgf \in [0,\,1]$,
we always have a trivial constraint
\begin{equation}
\fbf \le 1 - \faf\,.
\label{eqn:rbfconstraint1}
\end{equation}
Furthermore, we can rewrite $\fbf$ in a form
\begin{equation}
\fbf = \frac{4-\obf}{4-\oaf-\obf}\frac{3\obf}{4-\obf}\,.
\label{eqn:rbf1}
\end{equation}
Since $\frac{3\Omega_{\cb}}{4-\Omega_{\cb}}$ is an increasing function of time, and
$\fbs = \frac{3\obs}{4-\obs}$, we find
\begin{equation}
\fbf \le \frac{4-\obf}{4-\oaf-\obf}\fbs\,.
\end{equation}
The multiplier of $\fbs$ simplifies to $1+\faf/3$. So we have
a constraint
\begin{equation}
\fbf \le \left( 1 + \faf/3 \right) \fbs\,.
\label{eqn:rbfconstraint2}
\end{equation}
For example, when $\fbf=\fbs/2$ the constraint
(\ref{eqn:rbfconstraint1}) forbids a small region in the upper right
corner of Figs.~\ref{fig:ra} and \ref{fig:rb} where $\fbs \sim 1$
and $\faf \sim 1$.

Equality in Eq.~(\ref{eqn:rbfconstraint2}) corresponds to the case where $\obf=\obs$,
i.e., the second curvaton decays at the same moment as the first curvaton.
Then the transfer efficiencies simplify to $\ra=\faf$,
and $\rb=(1+\faf/3)\fbs=\fbf$, indicated by the thin solid lines
in Figs.~\ref{fig:ra} and \ref{fig:rb}.
(Note therefore that the case where $\fbf=\fbs$ describes a
situation where the curvaton $\cb$ decays some time {\em after} the curvaton $\ca$.)

\subsection{Quasi-second order }

In the next section we present the full second order calculation of $\zeta$,
but here we consider a simplified treatment, which will give the correct form of $\fNL$ generated by multiple curvaton decays in the limit where the non-Gaussianity is large.


In this quasi-non-linear approximation we will use the linearized form for the full curvature perturbation (\ref{eqn:zafullnl}) and~(\ref{eqn:zbfullnl}) in terms of the density perturbations on spatially flat hypersurfaces:
\begin{eqnarray}
\za & = & \frac{1}{3}\frac{\delta\rho_{\ca}}{\bar\rho_{\ca}}\,,
 \label{eqn:quasizetaa} \\
\zb & = & \frac{1}{3}\frac{\delta\rho_{\cb}}{\bar\rho_{\cb}}\,,
 \label{eqn:quasizetab}
\end{eqnarray}
but we will include non-linear contributions to the curvaton densities.
The energy density of the curvaton $\ca$ is $\rho_{\ca}\propto \ca^2=(\bar{\ca}+\delta\ca)^2$ and hence we have
\begin{eqnarray}
\frac{\delta\rho_{\ca}}{\bar\rho_{\ca}} & = &  2\frac{\delta\ca}{\bar{\ca}} + \left(\frac{\delta\ca}{\bar{\ca}}\right)^2\,.
\end{eqnarray}
%
For simplicity we neglect any non-linear evolution of the curvaton field [i.e., we take $g_{\ca}''=g_{\cb}''=0$ in Eq.~(\ref{defga}) and~(\ref{defgb})] so that the curvaton field perturbations can be taken to be Gaussian
and we have
\begin{eqnarray}
\za
     & = & \zal + \frac{3}{4}\zal^2\,,
\end{eqnarray}
where
\begin{eqnarray}
\zal & = & \frac{2}{3}\frac{\delta\ca_\ast}{\bar{\ca}_\ast}\,,
\end{eqnarray}
and similarly for the curvaton $\cb$.

Now we use the first order result for the primordial curvature
perturbation produced by the two curvaton decays, but instead of
$\zal$ and $\zbl$ we use the quasi-nonlinear $\za$ and $\zb$ of
Eqs.~(\ref{eqn:quasizetaa}) and~(\ref{eqn:quasizetab}). Then we have
from Eq.~(\ref{eqn:zgsl})
\begin{eqnarray}
   \zeta & = & \ra\za + \rb\zb \nonumber\\
  & = & \ra\left[\zal + \frac{3}{4} \zal^2 \right]
 + \rb \left[\zbl + \frac{3}{4} \zbl^2 \right]
 \,.
\end{eqnarray}
Hence, comparing this result with Eq.~(\ref{eqn:zetageneral}), we have in our quasi-second-order approximation
\begin{eqnarray}
A & = & \ra\,,\\
B & = & \rb\,,\\
C & = &  \frac{3}{2}\ra\,,\\
D & = &  \frac{3}{2}\rb\,,\\
E & = & 0\,.
\end{eqnarray}
Substituting these into Eq.~(\ref{eqn:fnl}) the nonlinearity parameter reads
\begin{equation}
f_{\rm NL}^{\rm quasi} = \frac{5}{4} \frac{ \ra^3
 + \beta^4 \rb^3}{\left(\ra^2+\beta^2 \rb^2\right)^2}\,.
\label{eqn:fnlquasi}
\end{equation}

We derived this result using only first-order formulas for $\zeta$,
but including the full second order expression for the energy
density. The true second order $\zeta$ includes corrections of order
$\zeta^2$, but we expect the above result to give a good
approximation when $\zn\gg\zl^2$, i.e., when the non-Gaussianity is
large.
In the next section we present the full second-order calculation.

\subsection{Second order}

\subsubsection{First decay}

At second order, Eq.~(\ref{eqn:nonlinf1}) reads
\begin{eqnarray}
\ogpre[4(\zgprel - \zfl)]^2 + \oaf [3(\zal - \zfl)]^2 + \obf [3(\zbl - \zfl)]^2 &&\nonumber\\
+ 4 \ogpre [\zgpren - \zfn] + 3 \oaf [\zan - \zfn] + 3 \obf [\zbn -
\zfn] & = & 0\,.
\end{eqnarray}
From this, the solution for the total curvature perturbation at the first decay is
\begin{eqnarray}
\zfn
& = & 4\fgf \left[\zgprel-\zfl\right]^2
 +3\faf \left[\zal-\zfl\right]^2
 +3\fbf \left[\zbl-\zfl\right]^2
 +\fgf\zgpren+\faf\zan+\fbf\zbn\,.
\label{eqn:zfn}
\end{eqnarray}
At second order, Eq.~(\ref{eqn:nonlinf2}) reads
\begin{equation}
\ogf[4(\zgfl - \zfl)]^2 + \obf[3(\zbl - \zfl)]^2
+ 4\ogf (\zgfn - \zfn) + 3\obf (\zbn - \zfn) = 0\,,
\end{equation}
which gives
\begin{eqnarray}
\zgfn
& = &  -4\left[\zgfl-\zfl\right]^2
 +3\left(1-\rf \right) \left[\zbl-\zfl\right]^2
 + \zfn +\left(1-\rf \right)\zbn -\left(1-\rf \right)\zfn\,.
\label{eqn:zgfn}
\end{eqnarray}

\subsubsection{Second decay}

We find the second order radiation perturbation after the second decay
from Eq.~(\ref{eqn:nonlins}) [or from (\ref{eqn:zgfn}) by first dropping
curvaton $\cb$ (which also implies $\rf=1$), then relabeling
$a\rightarrow b$, and $1\rightarrow 2$ and $0\rightarrow 1$]. The result is
\begin{equation}
\zgsn
= -4\left[\zgsl-\zsl\right]^2
+ \zsn\,.
\end{equation}
But recalling that $\zsl = \zgsl$ this is just
\begin{equation}
\zsn = \zgsn\,,
\end{equation}
exactly as it should be since after the second decay
only radiation is left.
Now $\zsn$ is found from (\ref{eqn:zfn}) by first dropping curvaton
$\cb$, and then relabeling $\ca\rightarrow\cb$, $1\rightarrow 2$
and $0\rightarrow 1$. Thus we have
\begin{eqnarray}
\zsn
& = & 4(1-\fbs)\left[\zgfl-\zsl\right]^2
 + 3\fbs\left[\zbl-\zsl\right]^2
 +(1-\fbs)\zgfn+\fbs\zbn\,.
\label{eqn:zgsn}
\end{eqnarray}
Here we substitute the curvature perturbations from our
calculations above and assume $\zgprel=\zgpren=0$. Then we end up  with
the second order part of the primordial curvature perturbation
\begin{equation}
\zn \equiv \zsn = \tilde C\zal^2 + \tilde D\zbl^2 + E\zal\zbl + F\zan + G\zbn\,,
\label{eqn:zgsnfinal}
\end{equation}
where
\begin{eqnarray}
\tilde C & = &
-2\rf^2\faf^2\fbs^2-\rf^2\faf^2\fbs^3+7\rf^2\faf^2\fbs-4\rf^2\faf^2\nonumber\\
&& +3\rf\faf^2-\faf^2-\rf\faf^3-\rf\fbf\faf^2+3\rf\faf-3\fbs\rf\faf^2\nonumber\\
&& +\fbs\faf^2+\fbs\rf\faf^3+\fbs\rf\fbf\faf^2-3\fbs\rf\faf  \label{eqn:Ccoeff}\\
\tilde D & = &
-1+\fbs\rf\faf\fbf^2-7\rf\fbf-\fbf^2+2\fbf+5\rf+\fbs-4\rf^2\fbf^2+8\rf^2\fbf\nonumber\\
&& +7\fbs\rf\fbf+7\rf^2\fbf^2\fbs-14\rf^2\fbf\fbs-2\fbs^2\rf^2\fbf^2+4\fbs^2\rf^2\fbf\nonumber\\
&& - \fbs^3\rf^2\fbf^2+2\fbs^3\rf^2\fbf-4\rf^2-5\fbs\rf+7\fbs\rf^2-2\fbs^2\rf^2-\fbs^3\rf^2\nonumber\\
&&
+3\rf\fbf^2-\rf\fbf^3+\fbs\fbf^2-2\fbs\fbf-\rf\faf\fbf^2-3\fbs\rf\fbf^2+\fbs\rf\fbf^3
  \label{eqn:Dcoeff}\\
E & = &
2\faf+2\fbs\rf\faf\fbf^2-2\rf\faf\fbf^2-10\rf\faf-2\rf\fbf\faf^2+10\fbs\rf\faf\nonumber\\
&& +2\fbs\rf\fbf\faf^2+14\rf^2\faf\fbs\fbf-4\fbs^2\rf^2\faf\fbf-2\fbs^3\rf^2\faf\fbf\nonumber\\
&& -6\fbs\rf\faf\fbf+8\rf^2\faf-8\rf^2\faf\fbf-14\rf^2\faf\fbs+4\fbs^2\rf^2\faf\nonumber\\
&& +2\fbs^3\rf^2\faf +6\rf\faf\fbf-2\faf\fbf-2\fbs\faf+2\fbs\faf\fbf   \label{eqnE:coeff}\\
F & = & \ra = \left(1-\fbs\right)\faf\rf \label{eqn:Fcoeff}\\
G & = & \rb = 1-\rf+\fbf\rf+\fbs\rf-\fbs\fbf\rf \label{eqn:Gcoeff}
\end{eqnarray}
In Appendix \ref{sect:appendixb} we rewrite these coefficients with $\rf$
from (\ref{eqn:rf}) substituted in and organized in order of raising
powers of $\faf$, $\fbs$, and $\fbf$.

Assuming linear evolution between Hubble exit and
the beginning of curvaton oscillation,
the genuine second order curvaton perturbations
in (\ref{eqn:zgsnfinal}) read
[see Eqs.~(\ref{eqn:zan}) and
(\ref{eqn:zbn})]
\begin{eqnarray}
\zan & = & -\frac{3}{2}\zal^2\,,\\
\zbn & = & -\frac{3}{2}\zbl^2\,.
\end{eqnarray}
Comparing now (\ref{eqn:zgsnfinal}) with (\ref{eqn:zetageneral})
we conclude
\begin{eqnarray}
C & = & \tilde C - \textstyle \frac{3}{2}F\,,\\
D & = & \tilde D - \textstyle \frac{3}{2}G\,.\\
\end{eqnarray}
Recalling that $A=\ra$ and $B=\rb$ from the first-order analysis, we now have found the non-linearity parameter $\fNL$ of Eq.~(\ref{eqn:fnl}):
\begin{equation}
\fNL = \frac{5}{6}
\frac{ \ra^2 \left(\tilde C - \textstyle \frac{3}{2}\ra \right) + \half
  \beta^2 E \ra \rb + \beta^4 \rb^2 \left( \tilde D - \textstyle \frac{3}{2} \rb \right)}
     {\left(\ra^{2}+\beta^2 \rb^{2}\right)^2}\,.
\label{eqn:fnlgen}
\end{equation}

\section{$\fNL$ in various limits}
\label{sect:variouslimits}

The general expression for $\fNL$ in Eq.~(\ref{eqn:fnlgen}) written
in terms of the coefficients $\tilde C$, $\tilde D$, $E$, $\ra$ and
$\rb$ in Eqs.~(\ref{eqn:Ccoeff})--(\ref{eqn:Gcoeff})
[or equivalently (\ref{eqn:b1})--(\ref{eqn:b2}) in Appendix B]
would be very complicated as a function of all four parameters
$\faf$, $\fbf$, $\fbs$ and $\beta$. Hence in this section we identify some
useful limiting cases in which the expression significantly simplifies.

\subsection{Single curvaton limits}

As a consistency check we first show that
if either of the curvaton densities is negligible when they decay then the result (\ref{eqn:fnlgen})
simplifies to the single curvaton decay result \cite{Sasaki:2006kq,LR,Bartolo}
\begin{equation}
  \fNLsingle(f) = \frac{5}{4f} - \frac{5}{3} -\frac{5f}{6}\,,
\label{eqn:fnlsingle}
\end{equation}
where
either $f=\faf$ if $\fbs=0$, or $f=\fbs$ if $\faf=0$.

If density of the second curvaton is always negligible, then we can
set $\fbf=\fbs=0$ which leads to
\begin{eqnarray}
\ra & = & \faf\,,\\
\tilde{C} & = & 3\faf - 2\faf^2  - \faf^3\,,
\end{eqnarray}
$\tilde D = E = \rb = 0$, and then Eq.~(\ref{eqn:fnlgen}) yields
\begin{equation}
 \label{singlera}
\fNL = \frac{5}{4\ra} - \frac{5}{3} -\frac{5\ra}{6} = \fNLsingle(\faf)\,.
\end{equation}

On the other hand, if the density of the first curvaton is
negligible, we have $\faf=0$ which leads to
\begin{eqnarray}
\rb & = & \fbs\,,\\
\tilde{D} & = & 3\fbs - 2\fbs^2  - \fbs^3\,,
\end{eqnarray}
$\tilde C = E = \ra = 0$, and then
\begin{equation}
\fNL = \frac{5}{4\rb} - \frac{5}{3} -\frac{5\rb}{6} = \fNLsingle(\fbs)\,.
\end{equation}
Hence in the limit where only one curvaton has a non-negligible
density, our two-curvaton result for $\fNL$ reduces to the
well-known single-curvaton result (\ref{eqn:fnlsingle}).

\subsection{Simultaneous decay of curvatons $\ca$ and $\cb$}

In the case where both curvatons decay at the same moment, we have
$\fbf=(1+\faf/3)\fbs$, as discussed after
Eq.~(\ref{eqn:rbfconstraint2}). Then we find
\begin{eqnarray}
\ra & = & \faf\\
\rb & = & \fbf\\
\tilde{C} & = & -\faf \left(\faf^2  + 2\faf + \fbf\faf - 3\right) \\
\tilde{D} & = & -\fbf \left(\fbf^2  + 2\fbf + \fbf\faf - 3\right)\\
E & = & -2\faf\fbf (\faf + \fbf + 2)\,.
\end{eqnarray}
The non-linearity parameter $\fNL$ which follows from these is
indicated by the thin solid lines in Figs.~\ref{fig:fnlbinflimit},
\ref{fig:fnlb1limit} and \ref{fig:fnlb0limit} for the cases
$\beta\rightarrow\infty$, $\beta=1$, and $\beta=0$ respectively.

In the limiting case where the second curvaton is homogeneous
($\beta=0$) the expression (\ref{eqn:fnlgen}) reduces to
\begin{equation}
 \fNL = \fNLsingle(\faf) - \frac{5}{6}\fbf\,.
\end{equation}
Conversely, in the limit when the first curvaton is homogeneous
($\beta\rightarrow\infty$) we get
\begin{equation}
 \fNL = \fNLsingle(\fbf) - \frac{5}{6}\faf\,.
\end{equation}
Note that in either case the example of two curvatons which decay at
the same time does not reduce exactly to the case of a single
curvaton. However
we can show that taking into account the constraint (\ref{eqn:rbfconstraint1})
the minimum value for $\fNL$ (for any value of $\beta$)
is still that found for a single curvaton, $\min(\fNL)=-5/4$.

\subsection{Both curvatons subdominant at decay}

If the energy density of both curvatons is small when they decay
then $\faf \ll 1$ and $\fbs\ll 1$ [which implies $\fbf \ll 1$, from
Eq.~(\ref{eqn:rbfconstraint2})]. To first order in the $f$
parameters, we find
\begin{eqnarray}
\ra & \simeq & \faf\\
\rb & \simeq & \fbs\\
\tilde{C} & \simeq & 3\faf\\
\tilde{D} & \simeq & 3\fbs\\
E & \simeq & 0\,.
\end{eqnarray}
In this case the non-linearity parameter Eq.~(\ref{eqn:fnlgen})
reduces to the quasi-second order result (\ref{eqn:fnlquasi}).
As expected the quasi-second order result gives a good approximation
when $\faf$ and $\fbs$ are small and the non-Gaussianity is thus
large.

\subsection{Second curvaton negligible at the first decay}
\label{sect:limitrbf0}

If the energy density of the second curvaton $\cb$ is negligible
when the first curvaton $\ca$ decays then we can set $\fbf=0$ and
the linear result reduces to the simpler result (\ref{Choilinear})
found by Choi and Gong \cite{Choi:2007fy}. We then have
\begin{eqnarray}
\ra & = & \big(1-\fbs\big)\faf\\
\rb & = & \fbs \\
\tilde{C}
& = & \faf \big(1-\fbs\big)\big( 3 -2\faf - \faf^2\big)
 + \faf^2\fbs \big(3 - 2\fbs - \fbs^2\big)\\
%
\tilde{D} & = & \fbs \big( 3-2\fbs-\fbs^2 \big) \\
E & = & -2 \faf \big( 3\fbs -2\fbs^2 - \fbs^3\big)\,
\end{eqnarray}
and
\begin{eqnarray}
\fNL & = & \frac{5}{6} \Big\{ (1-\fbs)^3\faf^3 \left[
\textstyle\frac{3}{2} - 2\faf - \faf^2 +
   \faf\fbs\left( \fbs+3 \right)\right]
 - \beta^2  (1-\fbs)^2\faf^2 \fbs^2 \left( \fbs+3
\right)\nonumber\\
&& + \beta^4 \fbs^3   \left( \textstyle\frac{3}{2} - 2\fbs - \fbs^2
\right) \Big\} \Big/ \left\{ (1-\fbs)^2\faf^2 + \beta^2\fbs^2
\right\}^2 \label{eqn:fnlrbf0}
\end{eqnarray}
In order to illustrate the general behavior of $\fNL$ we discuss
three particular cases.
\begin{figure*}[!th]
\includegraphics[width=0.72\textwidth]{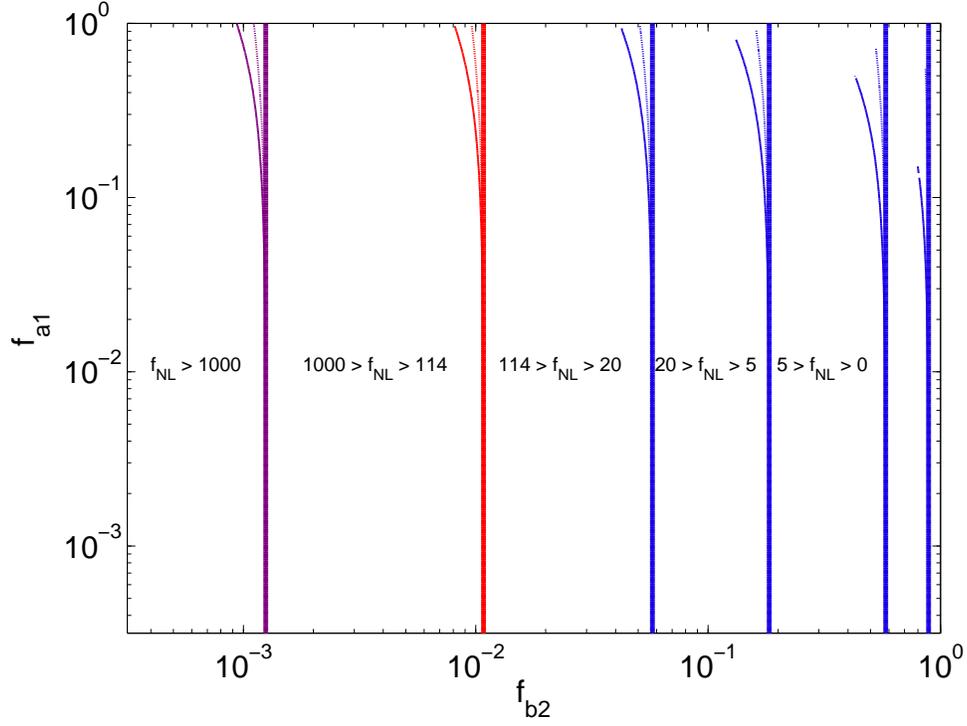}
\caption{$\fNL(\fbs,\faf)$ when $\beta\rightarrow\infty$ and $\fbf=0$ (\emph{thick
solid lines}) or $\fbf=\fbs/2$
  (\emph{dotted lines}) or $\fbf=(1+\faf/3)\fbs$ (\emph{thin solid lines}).
  In the latter two cases the constraint
  (\ref{eqn:rbfconstraint1}) excludes a small
  region in the top right corner.
  Contours of equal $\fNL$ are shown, from left to right, for $\fNL=$ $1000$, $114$, $20$,
  $5$, $0$, and $-1$.
\label{fig:fnlbinflimit}}
\end{figure*}
\begin{figure*}[!ht]
\includegraphics[width=0.72\textwidth]{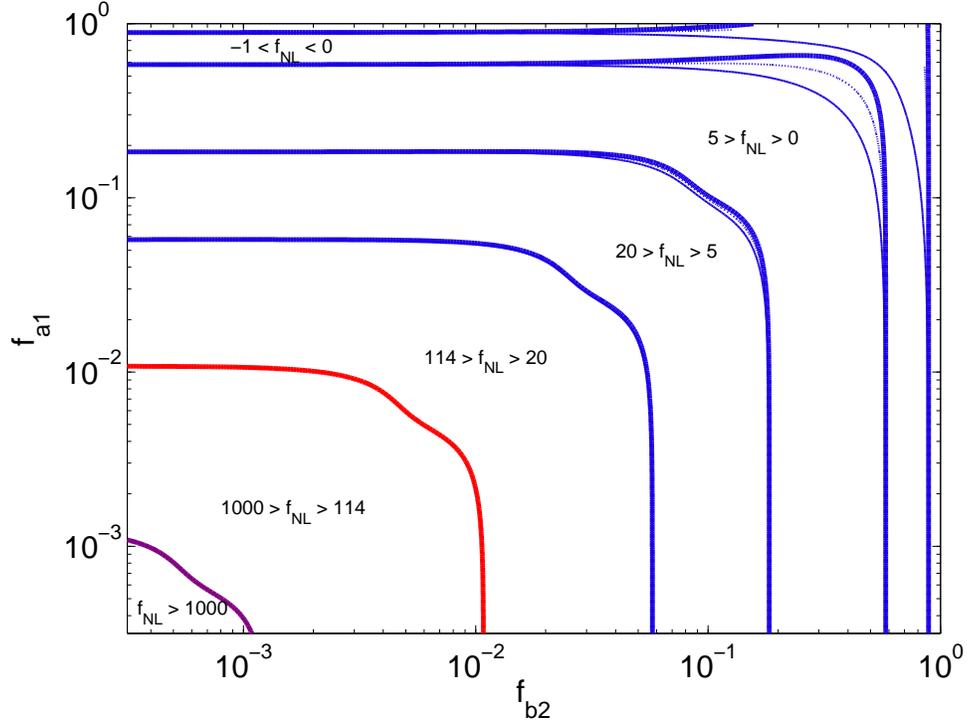}
\caption{Same as Fig.~\ref{fig:fnlbinflimit} but now $\fNL(\fbs,\faf)$
is shown for
  $\beta = 1$.
  Contours of equal $\fNL$ are shown, from the bottom left corner to the top right
  corner, for $\fNL=$ $1000$, $114$, $20$,
  $5$, $0$, and $-1$.
\label{fig:fnlb1limit}}
\end{figure*}
\begin{figure*}[!ht]
\includegraphics[width=0.72\textwidth]{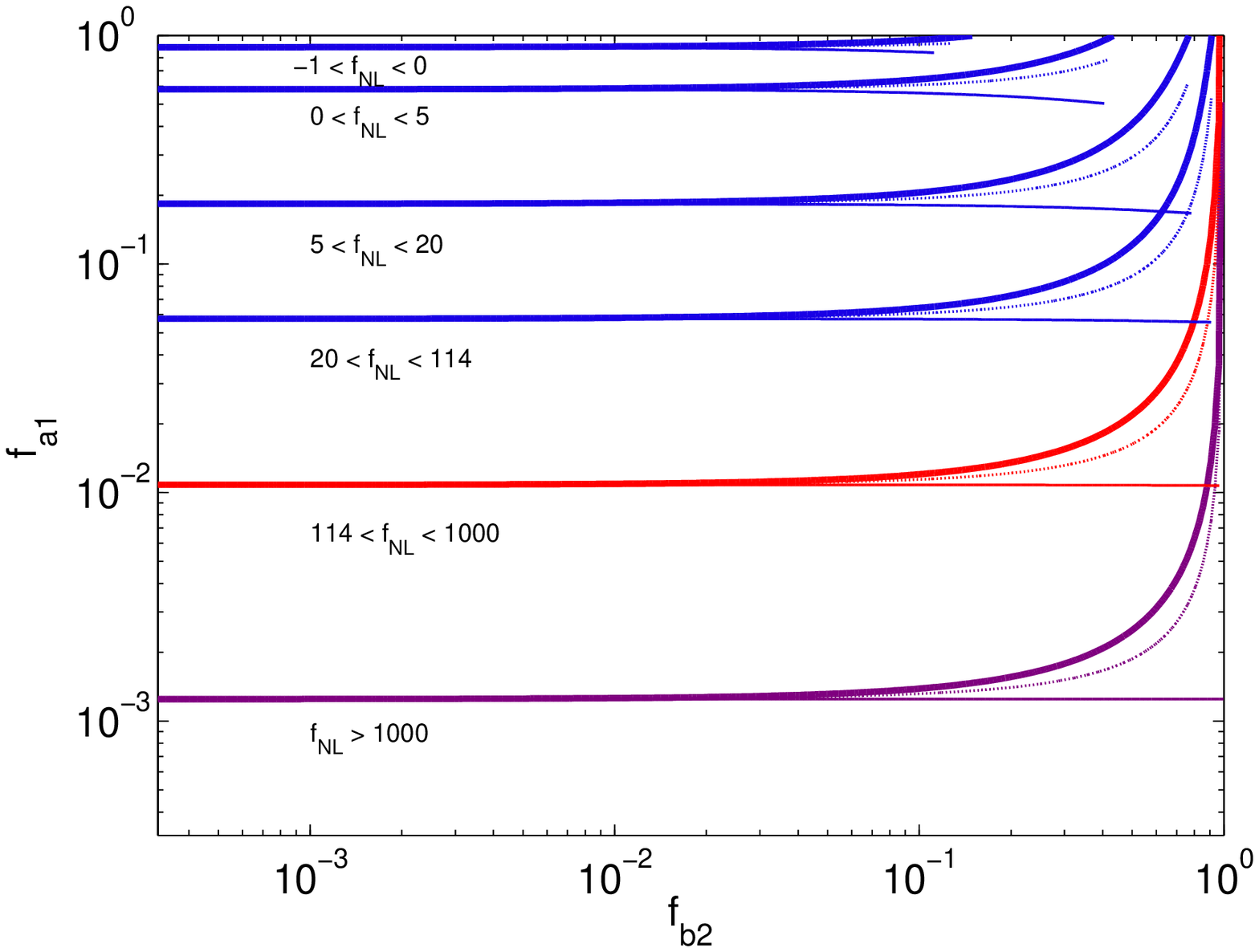}
\caption{Same as Fig.~\ref{fig:fnlbinflimit} but now $\fNL(\fbs,\faf)$
is shown for $\beta = 0$.
  Contours of equal $\fNL$ are shown, from bottom to top, for $\fNL=$ $1000$, $114$, $20$,
  $5$, $0$, and $-1$.
\label{fig:fnlb0limit}}
\end{figure*}


\paragraph{Case $\mathbf{\beta \rightarrow \infty}$:} this describes a
situation where the first curvaton is effectively homogeneous, $\za
= 0$. This case is shown by the thick solid vertical lines in
Fig.~\ref{fig:fnlbinflimit}.

In this case the universe remains homogeneous until the density of
the second curvaton becomes non-negligible. Thus there is no
difference from the original single inhomogeneous curvaton scenario,
and Eq.~(\ref{eqn:fnlrbf0}) gives
\begin{equation}
\fNL = \fNLsingle(\fbs)\,.
\end{equation}
Note that we have assumed here that the the density of second
curvaton is negligible when the first decays, $\fbf=0$. If this is
not the case then there can be a some dependence on $\faf$ (and
$\fbf$) illustrated in Fig.~\ref{fig:fnlbinflimit}.


\paragraph{Case $\mathbf{\beta = 1}$:} this illustrates a situation
where the fractional density perturbation in each curvaton field is
comparable, here $P_{\za} = P_{\zb}$. The thick solid lines in
Fig.~\ref{fig:fnlb1limit} show $\fNL(\fbs,\faf)$ corresponding to
Eq.~(\ref{eqn:fnlrbf0}) with $\beta=1$. We find that $\fNL$ can be
large in this case only when $\fbs\ll 1$ and $\faf \ll 1$
simultaneously.


\paragraph{Case $\mathbf{\beta = 0}$:} describes a situation where the
second curvaton is effectively homogeneous, ($\zb = 0$). With $\beta
= 0$ and $\fbf=0$ we have
\begin{equation}
\label{eqn:fnlb0limit}
 \fNL = \frac{1}{1-\fbs} \left[
\fNLsingle(\faf) + \frac{5}{6}\fbs \left(
  \fbs + 3 \right) \right]\,.
\end{equation}
In this case the inhomogeneous radiation produced by the first
curvaton decay is diluted by the decay of the second homogeneous
curvaton.
The thick solid lines in Fig.~\ref{fig:fnlb0limit} show $\fNL(\fbs,\faf)$
of Eq.~(\ref{eqn:fnlb0limit}).

When the second curvaton's density remains negligible throughout,
$\fbs \ll 1$, the non-linearity parameter is a function of a single
variable $\faf$, and we recover the single-curvaton case
(\ref{singlera}), which gives large $\fNL$ when $\faf \ll 1$.

In the opposite limit where $\fbs \rightarrow 1$, we find a new
regime in which the non-Gaussianity becomes large, which only
appears in the presence of two (or more) decaying scalar fields. The
non-linearity parameter is given by,
\begin{equation}
\fNL \rightarrow \frac{1}{1-\fbs} \left[ \fNLsingle(\faf) + \frac{10}{3}
\right]\,.
\end{equation}
When $\faf \rightarrow 1$, the expression in square brackets
approaches $25/12$.
Hence we have found an interesting result that $\fNL$ can be large
even when $\faf \sim 1$ if $\fbs \sim 1$.
Note that the quasi-second order approximation (\ref{eqn:fnlquasi}) correctly
predicts $\fNL \propto 1 / (1-\fbs)$ in this limit, but fails to
reproduce the numerical coefficient $25/12$.

\subsection{General case $\fbf \neq 0$}

As an example of the general case where the density of the second
curvaton is non-negligible when the first curvaton decays
($\fbf\neq0$), we show in Figs.~\ref{fig:fnlbinflimit}, \ref{fig:fnlb1limit} and
\ref{fig:fnlb0limit} $\fNL(\fbs,\faf)$ for the case where
$\fbf=\fbs/2$ and for the simultaneous decay where $\fbf=(1+\faf/3)\fbs$.
Generally we find only a weak dependence on $\fbf$, although the
constraint $\fbf \le 1-\faf$ excludes
a small region in the top right corner where $\fbs \sim 1$ and $\faf
\sim 1$.
For intermediate values of $\fbf\in [0, \,(1+\faf/3)\fbs]$ we find that the values of
$\fNL$ fall between the cases $\fbf=0$ and $\fbf=(1+\faf/3)\fbs$, i.e., the
contours of equal values of $\fNL$ in Figs.~\ref{fig:fnlbinflimit},
\ref{fig:fnlb1limit} and \ref{fig:fnlb0limit} lie between the
thick and thin solid lines shown.

Finally, we find numerically that the minimum value of the
nonlinearity parameter (\ref{eqn:fnlgen}) is $\min(\fNL) = -5/4$,
when the constraints (\ref{eqn:rbfconstraint1}) and
(\ref{eqn:rbfconstraint2}) are taken into account. This is the same
as the minimum value found for a single curvaton,
$\min(\fNLsingle)$, assuming that the evolution between Hubble exit
and beginning of curvaton oscillation is linear, which is the case,
for example, for a weakly interacting field with a quadratic
potential.

\section{Conclusion}
\label{sect:conc}

In this paper we have calculated the non-linear primordial curvature
perturbation, $\zeta$, following the decay of two curvaton fields in
the early universe. The full non-linear curvature perturbation is
given by the series of expressions, (\ref{eqn:nonlinf1}),
(\ref{eqn:nonlinf2}), (\ref{eqn:nonlins})
and~(\ref{eqn:nonlinfinal}), relating the final perturbation in the
radiation after the second curvaton decays, $\zgs$, to the initial
curvature perturbations in each curvaton field, $\za$ and $\zb$, in
the sudden-decay approximation. Expanding to first- and then
second-order in the perturbations we have derived the general
expressions in Eq.~(\ref{eqn:transfeffgeneral}) for the power
spectrum and Eq.~(\ref{eqn:fnlgen}) for the non-linearity parameter,
$\fNL$, which describes the non-Gaussianity of the primordial
curvature perturbation at leading order.

Large values of the non-linearity parameter in the two-curvaton model can qualitatively
be understood by a simplified quasi-second-order model
\begin{equation}
 \zeta \sim \ra \left[\zal + \frac34 \zal^2\right] + \rb \left[\zbl + \frac34 \zbl^2 \right] \,,
\end{equation}
leading to
\begin{equation}
 \fNL \sim \frac54 \frac{\ra^3+\beta^4\rb^3}{(\ra^2+\beta^2\rb^2)^2} \,
\end{equation}
where $\ra$ and $\rb$ are given by first-order results
(\ref{eqn:faformula}) and~(\ref{eqn:fbformula}).

We easily recover single field results: $\fNL\sim 5/(4\rb)$ if
$\faf=0$, or $\fNL\sim 5/(4\ra)$ if $\fbf=0$ and $\fbs=0$. In either
case the non-linearity becomes large if the transfer of curvaton to
curvature perturbation is inefficient ($\ra \ll 1$ and $\rb \ll 1$).
More generally in the two-curvaton model we find large $\fNL$ if
both curvatons are subdominant at their decay time.

We can also obtain large non-Gaussianity in a different way, only
possible due to the existence of two curvatons, where the second
curvaton is effectively homogeneous: $\beta=0$. In this case
$\fNL\propto 1/\ra$, but the efficiency $\ra\propto \faf(1-\fbs)$
becomes small either when $\faf$ is small (as in the one curvaton
case) or when $\fbs\sim 1$. In this latter case the inhomogeneous
radiation produced by the first curvaton decay is diluted by the
decay of the second homogeneous curvaton.

In all cases we find $\fNL\geq-5/4$, which seems to be a robust
lower bound in single and multi-field curvaton models in which the
curvaton field perturbations are themselves Gaussian, which should
be a good approximation for a weakly interacting field.


\begin{acknowledgments}
JV is supported by STFC and the Academy of Finland grant 120181, DW
by STFC, and HA by the Overseas Research Student Awards
Scheme (ORSAS).
\end{acknowledgments}

\appendix

\section{Notation}
\label{sect:appendixa}

The notation becomes quite complicated with the need to specify
two different curvaton species and the radiation produced in their decays,
which happen at two different moments. Furthermore, we need to keep track on
the first and second order parts of perturbations. Therefore, to advice the
reader, we present in this section a list of some symbols that appear
in our calculations and results.\\

%

\noindent A subscript $1$ or $2$ after specification of the particle
species tells whether the quantity is evaluated at first or second
decay. Note that some quantities are constant, and hence don't need this
subscript as apparent from the list below.\\

\noindent The energy density parameters ($\Omega_i = \rho_i / \rho_{\rm total}$):\\
$\oaf$ the density of the first curvaton $\ca$
just before the first decay,\\
 $\obf$ the density of the second curvaton $\cb$
at the first decay,\\
 $\obs$ the density of the second curvaton $\cb$ just before the second decay,\\
%
 $\ogpre$ the density of pre-existing radiation at the first decay,\\
 $\ogf$ the density of all radiation immediately after the first decay,\\
 $\ogfs$ the density of all radiation just before the second decay,\\
 $\ogs$ the density of all radiation immediately after the second decay.\\
%

\noindent The full non-linear curvature perturbations:\\
$\zeta$ ($=\zs$) the primordial perturbation after the second curvaton decay,
but before the nucleosynthesis,\\
$\zf$ the total perturbation at the first decay,\\
$\zs$ the total perturbation at/after the second decay,\\
%
%
$\za$ the perturbation of the first curvaton $\ca$,\\
$\zb$ the perturbation of the second curvaton $\cb$,\\
%
$\zgpre$ the pre-existing radiation perturbation,\\
$\zgf$ the radiation perturbation after the first decay,\\
$\zgs$ ($=\zs=\zeta$) the radiation perturbation after the second decay.\\

\noindent Subscript in parenthesis after all the other subscripts denotes the order
considered. For example:\\
$\zl$ the first order part of the primordial perturbation,\\
$\zn$ the second order part of the primordial perturbation,\\
$\zfl$ the first order part of the total perturbation at the first decay,\\
$\zfn$ the second order part of the total perturbation at the first decay,\\
$\zsl$ the first order part of the total perturbation at the second decay,\\
$\zsn$ the second order part of the total perturbation at the second decay,\\
$\zal$ the first order part of the first-curvaton perturbation,\\
$\zan$ the second order part of the first-curvaton perturbation,\\
$\zbl$ the first order part of the second-curvaton perturbation,\\
$\zbn$ the second order part of the second-curvaton perturbation,\\
$\zgprel$ the first order part of the pre-existing radiation perturbation,\\
$\zgpren$ the second order part of the pre-existing radiation perturbation,\\
$\zgfl$ the first order part of the radiation perturbation after the first decay,\\
$\zgfn$ the second order part of the radiation perturbation after the first decay,\\
$\zgsl$ the first order part of the radiation perturbation after the second decay,\\
$\zgsn$ the second order part of the radiation perturbation after the second decay.\\
%

%
%
%
%
%
%

\noindent The energy density ratios (curvature perturbation transfer
efficiencies) at the first decay:\\
$\faf = \fafexpr$,\\
$\fbf = \fbfexpr$,\\
$\fgf = \fgfexpr$,\\
$\rf = \rfexpr$.\\

\noindent The energy density ratios (curvature perturbation transfer
efficiencies) at the second decay:\\
$\fbs = \fbsexpr$,\\
$\fgs = \fgsexpr$.\\

\noindent The total first order curvaton perturbation transfer efficiencies:\\
$\ra = \frac{(1-\fbs)(3+\faf)\faf}{3(1-\fbf)+\faf}$,\\
$\rb = \frac{(1-\fbf)\fbs(3+\faf)+\fbf\faf}{3(1-\fbf)+\faf}$.\\

\noindent The non-linearity parameters:\\
$\fNLa$ the non-linearity parameter of the first curvaton perturbation $\za$,\\
$\fNLb$ the non-linearity parameter of the second curvaton perturbation $\zb$,\\
$\fNL$ the non-linearity parameter of the primordial perturbation $\zeta=\zs=\zgs$.\\


\section{Coefficients in the second order part of $\zeta$}
\label{sect:appendixb}

In this section we express the coefficients $\tilde C$, $\tilde D$, $E$, $F$
and $G$, Eqs.~(\ref{eqn:Ccoeff}-\ref{eqn:Gcoeff}), with $\rf$ from
Eq.~(\ref{eqn:rf}) substituted in. We don't write the neatest possible forms,
but instead organize the results in order of raising powers of
of $\faf$, $\fbs$, and $\fbf$. This is particularly practical for the
purposes of Sect.~\ref{sect:variouslimits}, where we express $\fNL$ in various limits. For example
in the limit where both curvaton are subdominant at their decay time, i.e.,
$\faf\ll1$, $\fbs\ll1$, and $\fbf\ll1$.

\newpage
The coefficients in the second order part of $\zeta$ in (\ref{eqn:zgsnfinal}) are:
\begin{eqnarray}
\tilde{C} & = & \frac{1}{(3-3\fbf+\faf)^2}\times\Big\{
\Big[27-27\fbf +(-27+27\fbf)\fbs\Big]\faf                        \nonumber\\
&& +\Big[-27\fbf+(27+27\fbf)\fbs-18\fbs^2-9\fbs^3\Big]\faf^2       \nonumber\\
&& +\Big[-18+3\fbf^2+(36-3\fbf^2)\fbs-12\fbs^2-6\fbs^3\Big]\faf^3  \nonumber\\
&& +\Big[-8+2\fbf+(11-2\fbf)\fbs-2\fbs^2-\fbs^3\Big]\faf^4          \nonumber\\
&& +\Big[-1+\fbs\Big]\faf^5
\Big\}\,,\label{eqn:b1}\\
\tilde{D} & = & \frac{1}{(3-3\fbf+\faf)^2}\times\Big\{
(27-54\fbf+27\fbf^2)\fbs+(-18+36\fbf-18\fbf^2)\fbs^2+(-9+18\fbf-9\fbf^2)\fbs^3 \nonumber\\
&&
+\Big[9\fbf-12\fbf^2+3\fbf^4+(18-45\fbf+30\fbf^2-3\fbf^4)\fbs+(-12+24\fbf-12\fbf^2)\fbs^2
\nonumber\\
&& \quad\quad\quad +(-6+12\fbf-6\fbf^2)\fbs^3
\Big]\faf \nonumber\\
&& +\Big[3\fbf-8\fbf^2+2\fbf^3+(3-9\fbf+11\fbf^2-2\fbf^3)\fbs+(-2+4\fbf-2\fbf^2)\fbs^2+(-1+2\fbf-\fbf^2)\fbs^3\Big]\faf^2 \nonumber\\
&& +\Big[-\fbf^2+\fbs\fbf^2\Big]\faf^3
\Big\}\,,\\
E & = & \frac{1}{(3-3\fbf+\faf)^2}\times\Big\{
\Big[18\fbf-18\fbf^2+(-54+36\fbf+18\fbf^2)\fbs+(36-36\fbf)\fbs^2+(18-18\fbf)\fbs^3\Big]\faf \nonumber\\
&&
+\Big[-24\fbf+6\fbf^3+(-36+60\fbf-6\fbf^3)\fbs+(24-24\fbf)\fbs^2+(12-12\fbf)\fbs^3\Big]\faf^2 \nonumber\\
&&
+\Big[-16\fbf+4\fbf^2+(-6+22\fbf-4\fbf^2)\fbs+(4-4\fbf)\fbs^2+(2-2\fbf)\fbs^3\Big]\faf^3 \nonumber\\
&& +\Big[-2\fbf+2\fbf\fbs\Big]\faf^4
\Big\}\,,\\
F & = & \ra = \frac{1}{(3-3\fbf+\faf)}\times\Big\{(3-3\fbs)\faf + (1 - \fbs)\faf^2
\Big\}\,,\\
G & = & \rb = \frac{1}{(3-3\fbf+\faf)}\times\Big\{(3 - 3\fbf)\fbs + \left[\fbf + (1-\fbf)\fbs\right]\faf
\Big\}\,\label{eqn:b2}.
\end{eqnarray}

\bibliography{MultiCurvatonNG}

\end{document}